\documentclass[11pt]{article}

\usepackage{graphicx}

\usepackage{amsmath}
\usepackage{amssymb}
\usepackage{amsxtra}

\title{The multicomponent dark matter structure and its possible observed manifestations}
\author{V. Beylin\\Southern Federal University, Institute of Physics\\ vitbeylin@gmail.com\\ \fbox{V. Kuksa}\\Southern Federal University\\ M. Bezuglov\\ Laboratory of Theoretical Physics, JINR, Dubna\\ bezuglov.ma@phystech.edu\\ D. Sopin\\ Southern Federal University, Physical Department\\ sopindo@mail.ru}

\begin{document}
\maketitle

\begin{abstract}
In the framework of hypercolor extension of the Standard Model having vectorlike hyperquarks and two stable dark matter candidates originated from different hyper-currents, we consider some effects which result from reactions with participation of the dark matter components. Namely, there are decays of charged hyperpions into leptons and neutral component, annihilation and transitions of heavy dark matter candidates into the light ones. In the last case, low energy photon radiation from intermediate charged states is possible. This type of the dark matter luminescence is analyzed in more detail.  
\end{abstract}



\section{Introduction}

Experimental evidence of symmetry breaking mechanism in the Standard Model (SM) and the existence of Higgs bosons with the properties predicted have even worsened, in some sense, the situation in high-energy physics. From this time the SM has acquired the features of a complete and closed theory, without being such. This obviously means that the SM is only a kind of limit of a more general theory that should give solutions of open problems of the SM. Unfortunately, the inspiring idea of supersymmetry (SuSy) does not manifest itself in experiments at a scale of $\sim 1\, \mbox{TeV}$, which reduces the potential of the theory, although it does not completely close it.
Then, the search for ideas to solve the SM problems leads to analysis of the extension options for the SM. A lot of variants of the generalization and extension of the SM have been discussed in literature, in particular, those that offer different ways of explaining the structure and properties of dark matter (DM). At the moment, the DM existence is a firmly established fact, which is confirmed by many astrophysical observations. About a quarter of the universe mass is the DM and it plays a crucial role in the evolution of galaxies. Despite the fact that the existence of this substance has been known for an almost a century, and that it literally surrounds us, we still do not have the faintest idea what it is. All that we have is a set of hypothetical essences for its explanation, they are neutralinos from SuSy, axions, sterile neutrinos, inert Higgses, primordial black holes, manifestations of modified dynamics and last but not list WIMPs.  The most popular option is considered to be the WIMPs and they will be discussed in this work. It should be added that all efforts to catch the DM particles directly do nor successful up to now, and indirect methods to see some and measure any signals of processes with the DM participation become much more important\cite{Revv,Roszkowski,Indirect1,Khlopov1,Indirect2,DMsearch,IndirectB}.

Multicomponent models of the DM have become the object of attention and study mainly in recent decade\cite{Mcdm,Rev1} because, on the one hand, various variants of the SM extension were proposed suggesting some possible types of the DM carriers. On the other hand, there are unexplained astrophysical phenomena that can be better interpreted and explained within the framework of a multicomponent DM scenario. These phenomena are, in particular, monochromatic photon signals of unknown origin from Galaxy center with energies up to tens GeV\cite{GCexcess,Hess} and some features of spectra of cosmic leptons (positron excess, for instance)\cite{Fermi,Ams}.

In accordance with these two aspects, it is possible to divide the proposed DM variants into some two classes. In the first one, a multicomponent DM is resulted from a specific symmetry, which extend and generalize the SM group of symmetry and consequently introduce some additional degrees of freedom providing stability for part of them. What manifestations and specific effects can be induced by these new particles - it depends on the properties of the model symmetry and interactions. But the second type of scenarios introduces new Dark Matter candidates aiming the explanation of the observed physical phenomena, for example, the positron fraction excess in cosmic-ray spectra (leptophilic models \cite{Zurek,DMPositrons}) or to interpret photon signals (gamma emission from Galactic center region) as the result of annihilation of DM particles. Certainly, to stabilize new objects the initial SM symmetry should be also modified, for instance, using discrete symmetries. Namely, imposing a ZN symmetry (it can results from spontaneously broken U(1)) provides simultaneously an existence of several stable scalar fields as the DM candidates \cite{Yaguna}.

As another example, the renormalizable extension of the SM by a scalar, pseudoscalar and a singlet fermion fields are considered  where the DM has a fermion and a scalar components \cite{Majumdar,Majumdar1}. It allows to explain  photon signal with energy in a keV region by the light scalar decay, and 130-GeV photons emerging, for example, as a consequence of heavy fermions annihilation.
Another approach to introduce and use a two-component Dark Matter is to add a neutral Majorana fermion and a neutral scalar singlet interacting with the SM fields through the Higgs portal. Fermion, however, interacts at the tree level as Yukawa particle. And again, in various regions for the mass of the scalar photonic signal can be interpreted as result of of the scalars annihilation\cite{Drozd}. 

As most obvious case, some co-existence of axions with neutralino or wino\cite{AxionN,AxionW} allows to keep the SuSy scale near $1\, \mbox{TeV}$ (however, it is difficult to provide the necessary value of the DM relic density). There also suggested also an interesting way to use Exceptional Supersymmetric Standard Model (E6SSM) \cite{Khalil}, where two DM components arise from the set of Higgs superfields due to discrete symmetries again.

The DM candidates can be built from any suitable "matter" - additional scalars, fermions, even strongly interacted objects, Higgses portal, dark atoms, the DM can be presented by elementary particles or some compound states. In any case, these DM candidates should be neutral and stable, and channels and steps of their production in direct experiments at colliders together with their indirect manifestations in astrophysics phenomena or observed by space telescopes and at ground observatories (IceCube, LHAASO etc.) are analyzed carefully in a lot of papers\cite{Bhattacharya1,Bhattacharya2,Mambrini,Dasgupta,Grzadkowski,MinScalar,ScalarFermion,Bhattacharya3} and also in\cite{Bll,Bll1,Higgses1,Higgses2,Higgses3,InertVect,ScalarVect,SterileN,Multi_N,Kuksa1,DarkA,HadRev}. Certainly, all possible scenarios consider two aspects of the DM physics: theoretical validity and self-consistency of the model, and (qualitative and quantitative) description of the observed specific effects.  

An existence of several DM components substantially increases the number of reactions with the DM participation and allow to predict some interesting channels of its manifestations. The most important for such predictions is the structure of the DM sector in the model and features of the dark matter interactions with the SM particles and between the DM components. To clarify this possibility, we consider hypercolor model with additional heavy fermions (hyperquarks) in confinement, which can produce a set of composite states, hyperhadrons, in the framework of $\sigma-$model at some high scale\cite{Sundrum,Antipin1,Ourprd,Weadv} in an analogy with low-energy quark-meson theory. So, in the model 
a number of pseudo-Nambu-Goldstone (pNG) particles emerges, they acquire masses after the chiral symmetry breaking. There arise fifteen pNG states (and their chiral partners) which are connected with corresponding H-quark currents. The model includes almost standard Higgs boson which is (slightly) mixed with scalar $\sigma-$ meson.

Specifically, the model contains several neutral stable particles they can be interpreted as the DM candidates. In more detail, these states and their main characteristics will be presented in Section 2. Section 3 will be devoted to analysis of some new effect induced by the complex structure of the DM sector - radiation of photons in transitions of one of the DM component into another one. In Conclusion we discuss some possible application of this effect.

\section{Hyperquark scenario: features of the dark matter sector}							

The SM content and possibilities can be extended by introducing a new fermion sector in confinement using, for example, an extra gauge symmetry $SU (2) _ {tc}$. Besides, an additional $SU (2) _w$ symmetry should be to ensure electroweak interaction of new fermions (H-quarks) with the SM fields. Then, the hypercolor model in its minimal form contains only one doublet (with zero hypercharge) of heavy Dirac H-quarks doublet and keeps the standard Higgs boson. It, however, will mix with scalar $\tilde \sigma-$ meson generated by extra singlet scalar field, which is necessary for a spontaneous symmetry breaking. As a result, the new fields can acquire masses. Note, the mixing between scalars should be small to ensure the stability of precisely measured SM parameters (in other words, oblique corrections of Peskin Tackeuchi should be sufficiently small). 

In an analogy with the low-energy hadron QCD-based theory, H-quarks should form H-hadrons, which can be described in the H-$\sigma-$ model with an effective Lagrangian. So, there arises (due to global SO(4) symmetry breking) a set of pseudo-Nambu-Goldstone (pNG) states: a triplet of pseudoscalar H-pions and one neutral H-baryon along with its antiparticle. More exactly, H-baryon is a diquark state having an additive conserved quantum number, H-pions possess a multiplicative conserving quantum number\cite{Bai,Weadv}. Importantly, neutral states, $tilde \pi^0$ and $B^0,\, \bar B^0$, are stable. Consequently, they can be interpreted as the DM candidates with equal masses at the tree level. 

Because we are mostly interested to analyze the stable candidates properties, we do not consider heavier unstable H-hadrons and H-mesons
(see, however, study of their mass spectrum at lattice \cite{Lat1,Lat2} in the same gauge $Sp(4)$ theory).  Then, we need to know tree-level masses of lowest states (H-pions and H-baryon), mass of H-sigma and its v.e.v.; all of them are supposed to be $O(TeV)$. The angle of mixing, $\theta$, between  H-sigma and the Higgs boson, as it dictates by Peskin - Tackeuchi parameters for the model, should be such that $\sin \theta \lesssim 0.1$. 				

There were calculated both the mass splitting between neutral and charged states in H-pion triplet (induced by electroweak loops only) and between the lowest states of different origin (H-pions and H-baryons). In the last case, the mass splitting depends on some renormalization scale because of different H-quark currents generating these H-states, so, the mass splitting can be as much as tens of GeV. Electroweak mass splitting in H-pions triplet is well known and it is $\approx 160$ GeV. So, in this minimal scenario there are three stable particles possibly constituting dark matter: neutral stable H-baryon along with its antiparticle (we will consider them as the one component) and the lightest neutral H-pion.
	
To estimate their masses, there was used known way of analysis of the dark matter density evolution to its modern value. Namely there were written down five equations for each DM component taking into account charged H-pions that are decayed eventually into neutral one, i.e. so-called, co-annihilation processes were also considered for H-baryons and H-pions.
Numerical solution of the system of equations demonstrates that the correct value of the DM abundant corresponds to some areas in the parameter plane-of H-pions and H-sigma mases. Despite of the DM candidates masses estimation (approximately, they are in the region $0.8 - 1.2$ TeV), it was found that in all permitted areas of parameters $B^0-$ baryons dominate in the DM density\cite{IJMPA_2019}.
The reason of this asymmetry for the DM components contributions into the total density follows asymmetry of their interactions with the SM matter: H-pions have EW channels, but H-baryons do not participate in tree level EW interactions, they use (pseudo)scalar exchanges through Higgs boson and/or $\tilde \sigma-$meson instead. It is an important consequence of different origin for these DM components providing slower burnout of $B^0$ component in comparison with H-pion component.

Remind, both DM components were considered initially as having equal masses, mass splitting in the H-pion triplet is defined only EW loops and it is small. However, one-loop mass splitting between $\tilde \pi^0$ and $B^0$ can be as high as $10 - 15$ TeV depending on $\tilde \sigma-$ meson mass and value of renormalization parameter, $\mu$. Again, it due to the different structures of H- quark currents with which these components are associated. Corresponding mass splittings are demonstrated in Fig.1a,b. Note, mass of $\tilde \sigma-$ meson is near the value which is dictated by relation $m_{\tilde \sigma}^2\approx 3\cdot m_{\tilde \pi}^2$ resulted from zero $H-\tilde \sigma$ mixing.

 \begin{figure}[h]
 	\begin{minipage}[h]{0.5\linewidth}
 		\centering{\includegraphics[width=0.7\linewidth]{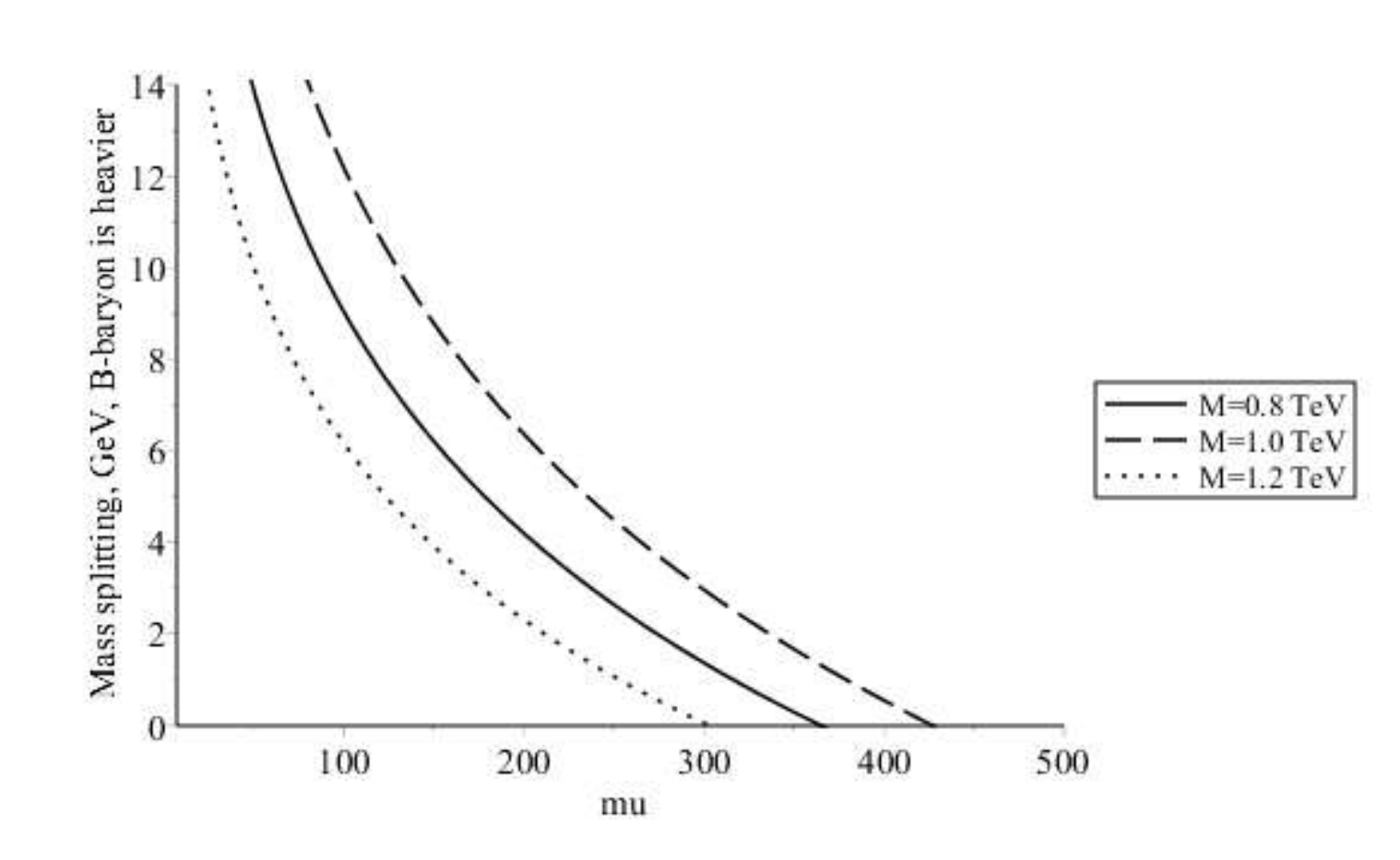}\\a)}
 	\end{minipage}
 	\hfill
 	\begin{minipage}[h]{0.5\linewidth}
 		\centering{\includegraphics[width=0.7\linewidth]{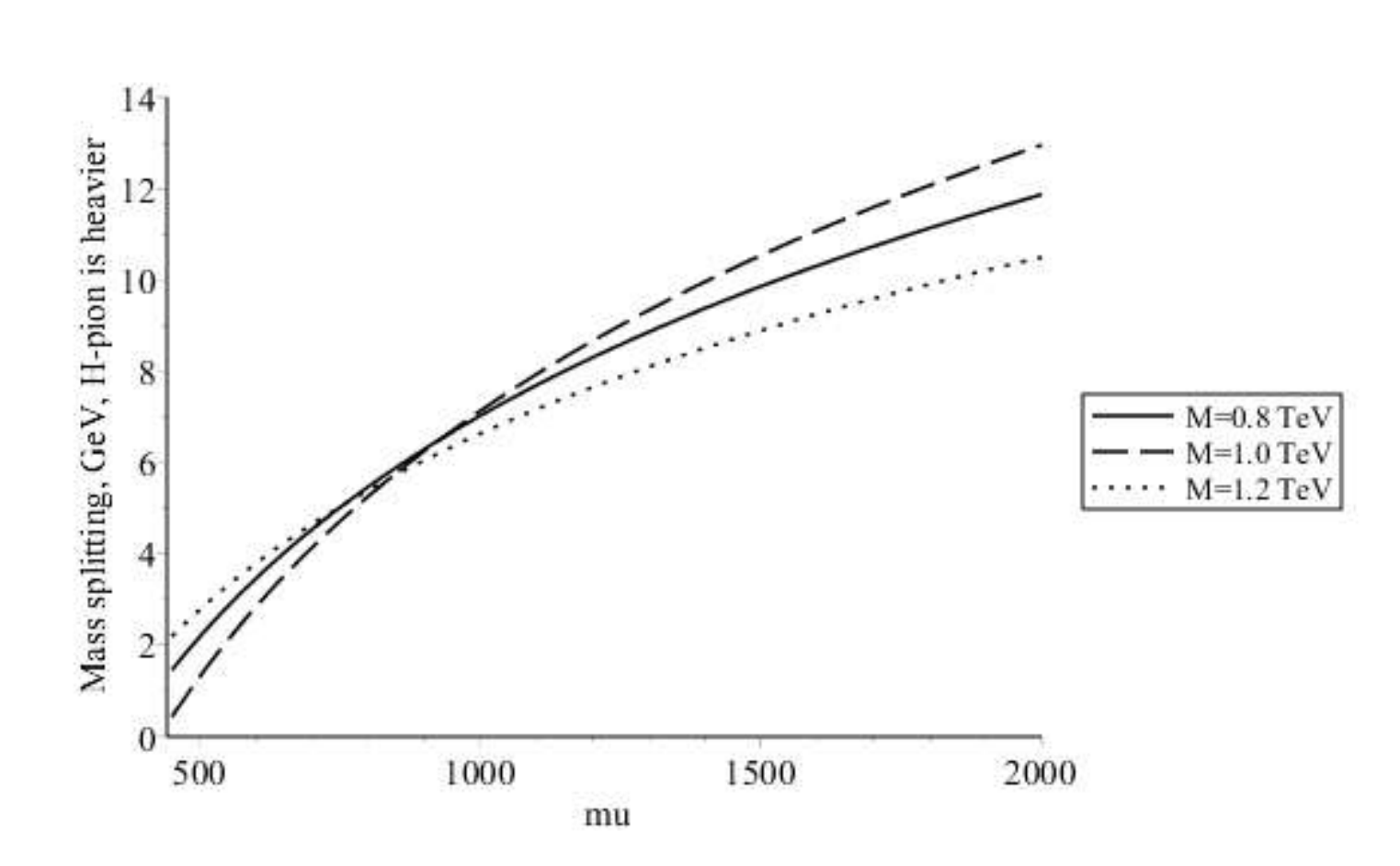}\\b0}
 	\end{minipage}
 	\vfill
  	\caption{Mass splitting between DM components in dependence on renormalization scale: a) $B^0$ is heavier; b) $\tilde \pi^0$ is heavier.}  \label{Delta}
 \end{figure}
 
 For nonzero mass splitting between the components, there occur an interesting process of the heavier DM component transformations into the lighter one. It can result to some effects, which are specific for suitable scenarios of multicomponent DM. Here, we will consider the case when $B^0$ is heavier than $\tilde \pi^0$, it is that the scenario when the tree level process of annihilation of heavy $B^0B^0$ pair into H-pions can be accompanied with some final state radiation (FSR).  
 
 \section{Transitions between DM components and an effect of luminescence}

To discuss possible $\gamma-$radiation in the transitions between dark matter components, we firstly analyzed the ratio of cross section of $B^0B^0$ pair annihilation into H-pions (see diagrams in Fig.2a) to the total cross section of $B^0B^0$ pair annihilation into all possible SM final states.

\begin{figure}[h]
	\begin{minipage}[h]{0.7\linewidth}
		\centering{\includegraphics[width=0.7\linewidth]{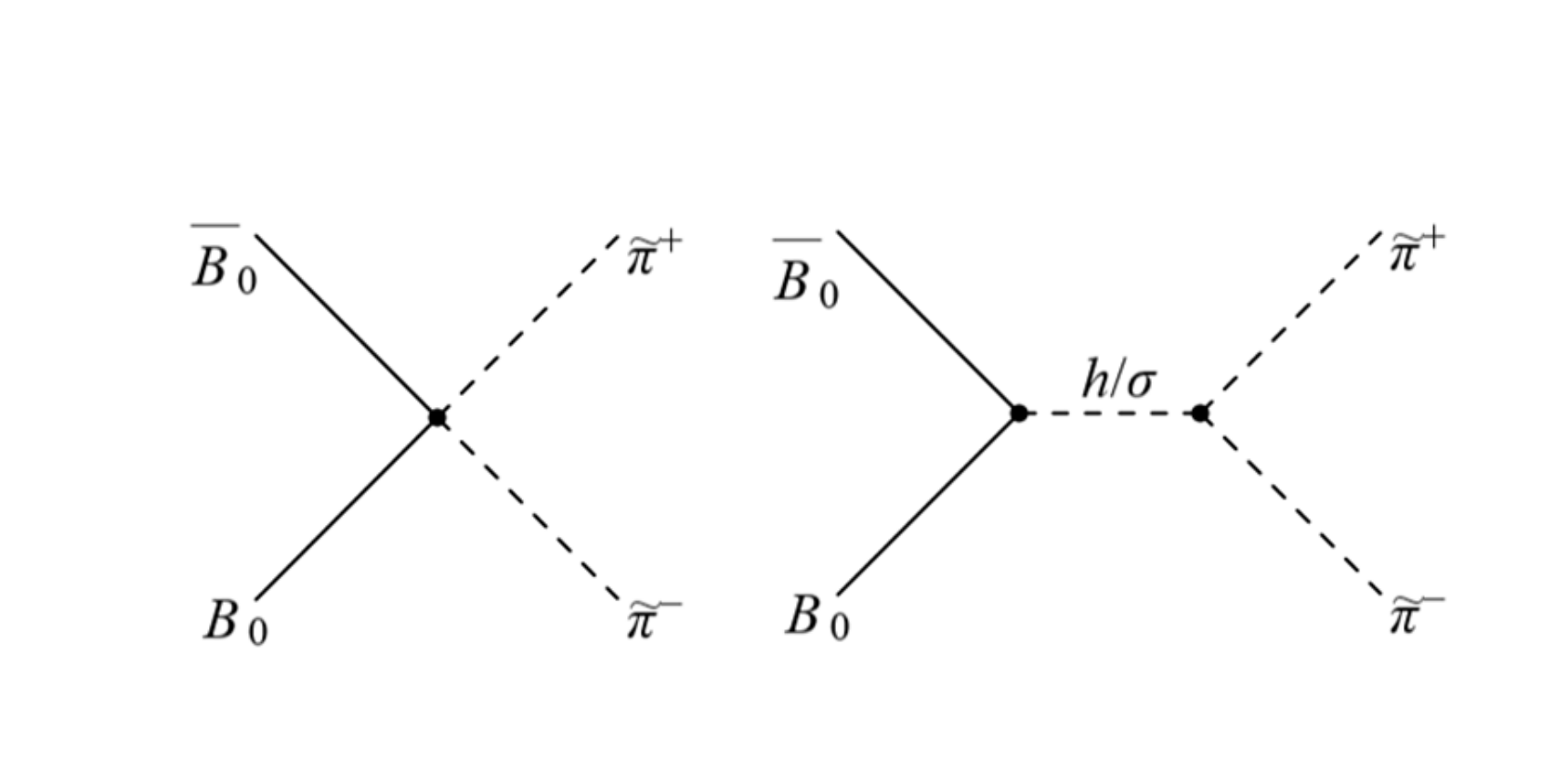} \\ a)}
	\end{minipage}
	\vfill
	\begin{minipage}[h]{0.8\linewidth}
		\centering{\includegraphics[width=0.8\linewidth]{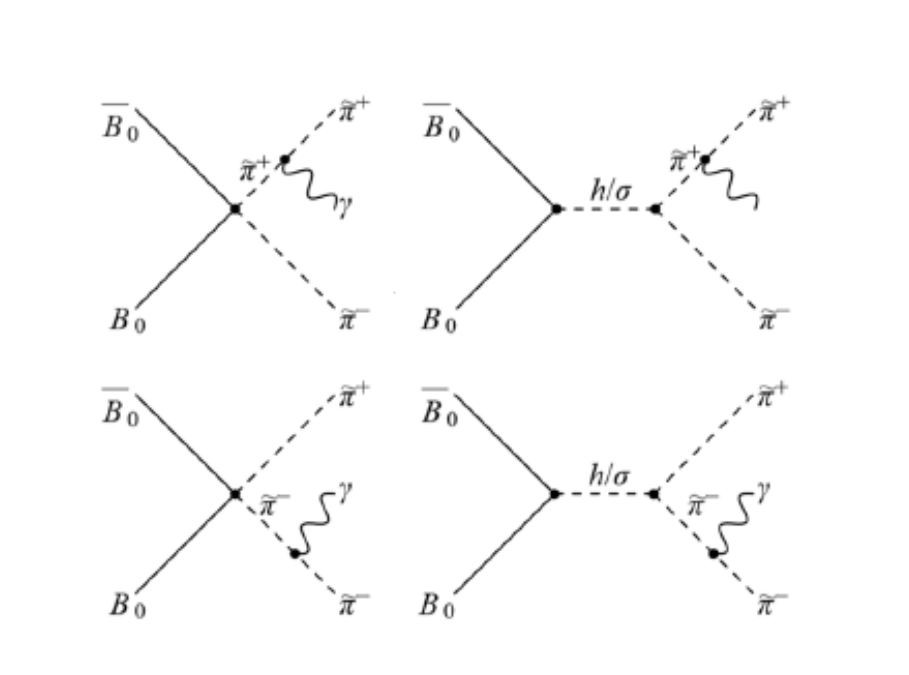} \\ b)}
	\end{minipage}
	\caption{Diagrams for $B^0$ annihilation into H-pions: a) without FSR, b) with FSR.}
	\label{Diags}
\end{figure}

Denoting this ratio as $\alpha$, we consider its values in the $\tilde-\pi - \tilde \sigma$ plane for various sets of model parameters: scale of renormalization (it also determines the mass splitting between $B^0$ and $\tilde \pi$ states), mixing angle, and the vacuum shift for a heavy scalar field. Some regions of $\alpha$ values are shown in Fig.3;  in all cases it is possible to find an areas where  $\alpha$ parameter is sufficiently large, $\alpha\geq 10$. Fortunately, in these regions H-pions and $B^0$-H-baryons masses are $\sim 1$ TeV, as it also follows from kinetics of the DM burning out. Thus, it is possible to fix a suitable interval of the DM components and the sigma meson masses, at which the $BB-$pair transition into charged unstable H-pions dominates.

\begin{figure}[h]
	\centering{\includegraphics[width=0.5\linewidth]{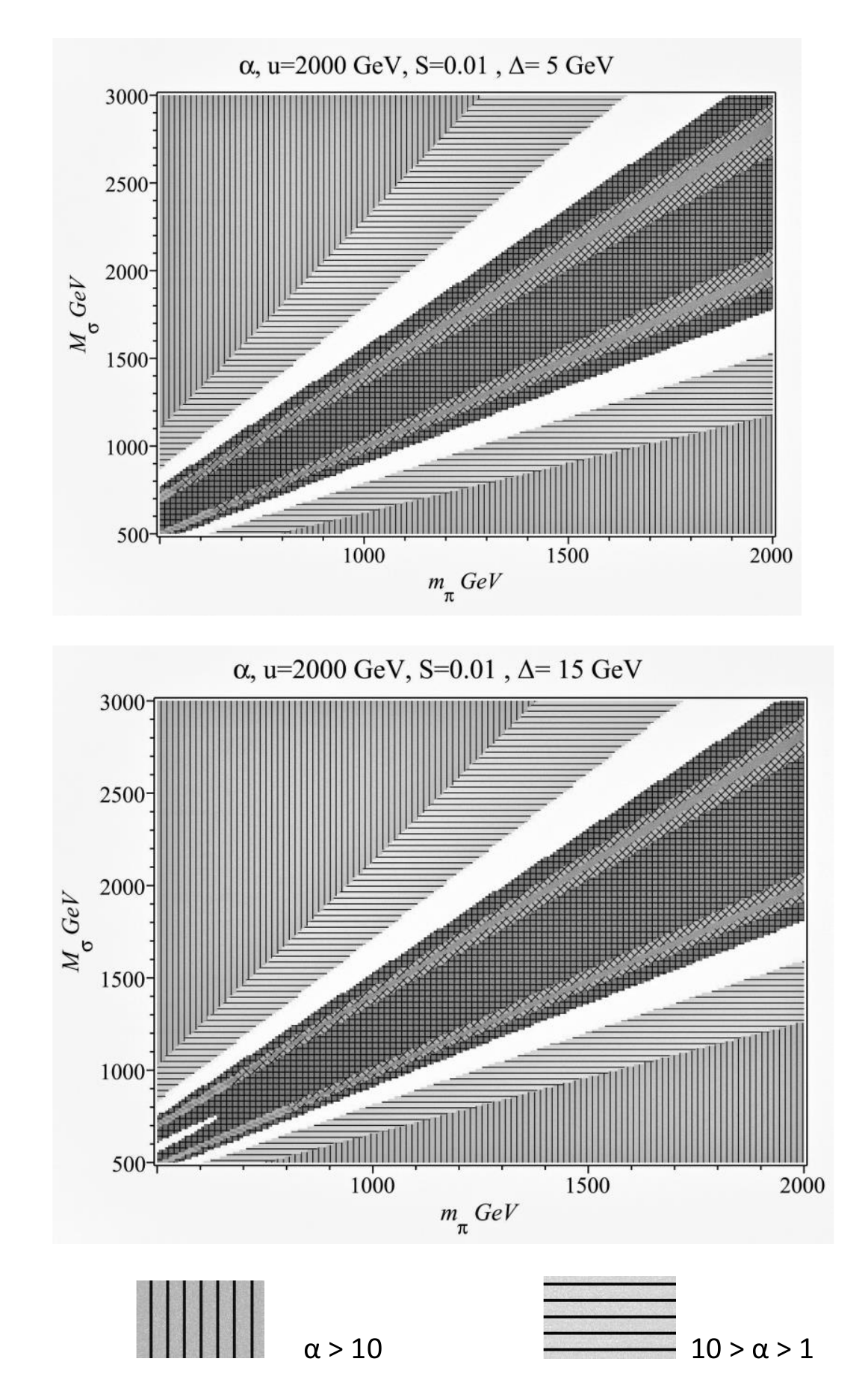}}
	\caption{Ratio of the $B^0B^0$ annihilation cross sections, notations of important regions for parameter $\alpha$ are depicted.}
	\vfill
	\label{Regs}
\end{figure}

Effect of FCR occurs just in the reaction $B^0B^0 \to \tilde \pi^+ \tilde \pi^- + \gamma$ (see diagrams in Fig. 2b) with subsequent decays of charged H-pions, $\tilde \pi^+\to \tilde \pi^0 +l \nu_l$. These charged states decay through strong and EW channels producing neutral stable $\tilde \pi^0$ and pair of lepton plus (anti)neutrino; corresponding widths \cite{Weadv} are: 

\begin{align}\label{Width}
\Gamma(\tilde{\pi}^{\pm}\to\tilde{\pi}^0 \pi^{\pm})&=6\cdot
10^{-17}\,\mbox{GeV},\,\,\,\tau_{\pi}=1.1\cdot10^{-8}\,\mbox{s},\,\,\,c\tau_{\pi}\approx 330\,\mbox{cm};\notag\\
\Gamma(\tilde{\pi}^{\pm}\to\tilde{\pi}^0 l^{\pm}\nu_l)&=3\cdot10^{-15}\,\mbox{GeV},\,\,\,\tau_l=2.2\cdot10^{-10}\,\mbox{s},\,\,\,c\tau_l\approx 6.6\,\mbox{cm}.
\end{align}
Now, for the differential cross section we get the following expression:

\begin{gather}\label{dsdE}
\frac{d\sigma v(B^0B^0 \to \tilde \pi^+ \tilde \pi^- \gamma)}{dE_{\gamma}} = \frac{4\alpha_e \sigma v(B^0B^0 \to \tilde \pi^+ \tilde \pi^-)}{\pi M_BE_{\gamma}\sqrt{M_B^2 - m_{\tilde\pi}^2}}\cdot\\ \nonumber \Bigl(-2\sqrt{M_B(M_B-E_{\gamma})}\sqrt{M_B(M_B-E_{\gamma})-m_{\tilde\pi}^2}+\\ \nonumber (2M_B(E_{\gamma}-M_B)+m_{\tilde\pi}^2)\log [\frac{2\sqrt{M_B(M_B-E_{\gamma})}}{\sqrt{M_B(M_B-E_{\gamma})}+\sqrt{M_B(M_B-E_{\gamma})-m_{\tilde\pi}^2}}-1]\Bigr ).
\end{gather}

So, possibility of radiation from (unstable) charged components (of H-pion triplet) is a specifics of the SM extensions with a complex structure resulting to the multi-component DM. If there are suitable channels of interaction, heavier DM component can transform into the light one, but for the FSR (or virtual internal bremsstrahlung) occurrence this transition should have an intermediate stage with some charged states. Here is exactly the same case.

Cross sections for different values of the DM component masses, mixing angle, mass of $\tilde \sigma-$meson and scale of H-symmetry breaking are presented in Fig.5 and 6. Here are shown also total cross sections demonstrating an obvious s-channel resonance near $M_{\tilde \sigma}$ in Fig. 7.
. 

\begin{figure}[h]
	\begin{minipage}[h]{0.5\linewidth}
		\centering{\includegraphics[width=0.7\linewidth]{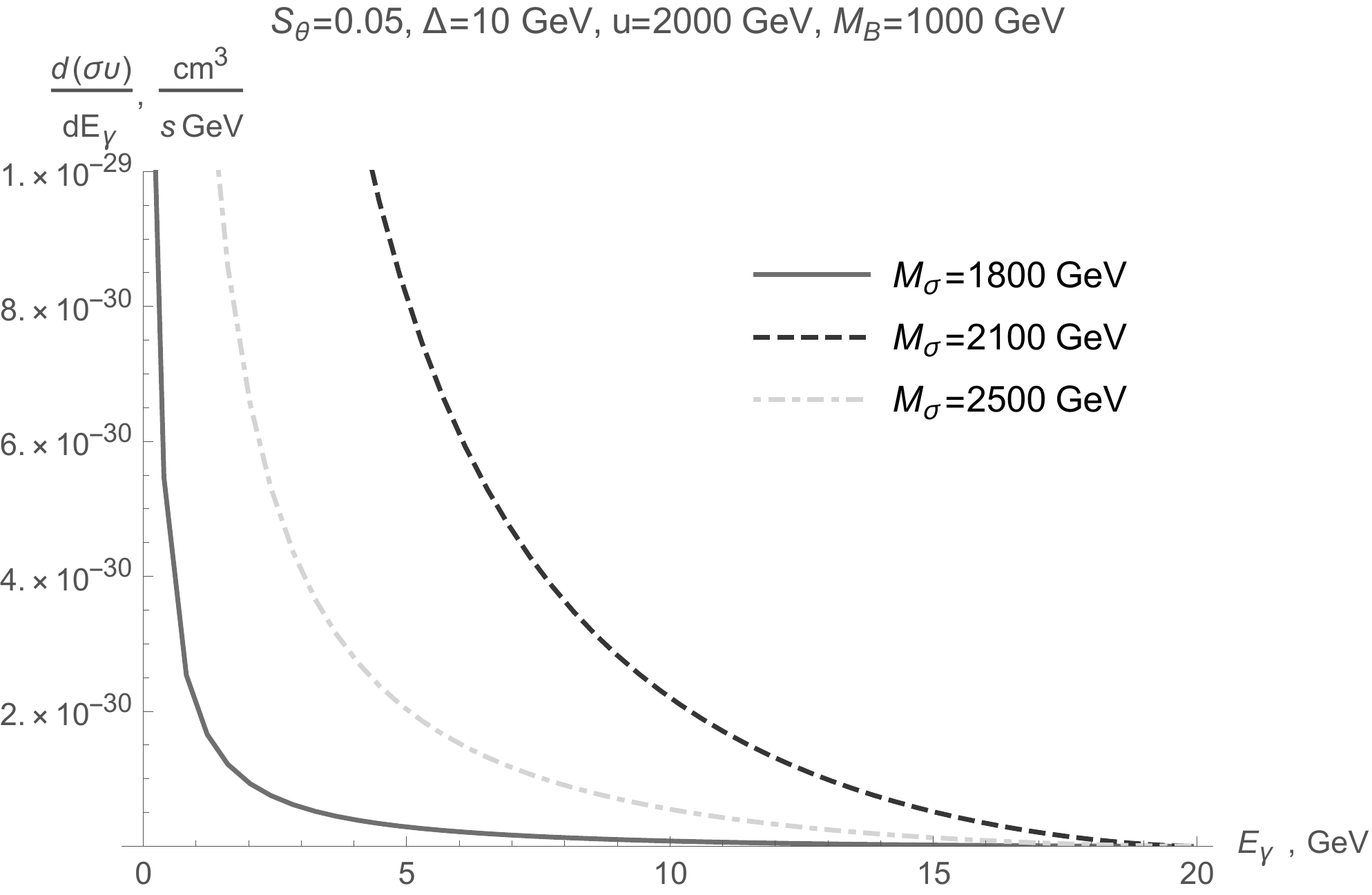}}
	\end{minipage}
	\hfill
	\begin{minipage}[h]{0.5\linewidth}
		\centering{\includegraphics[width=0.7\linewidth]{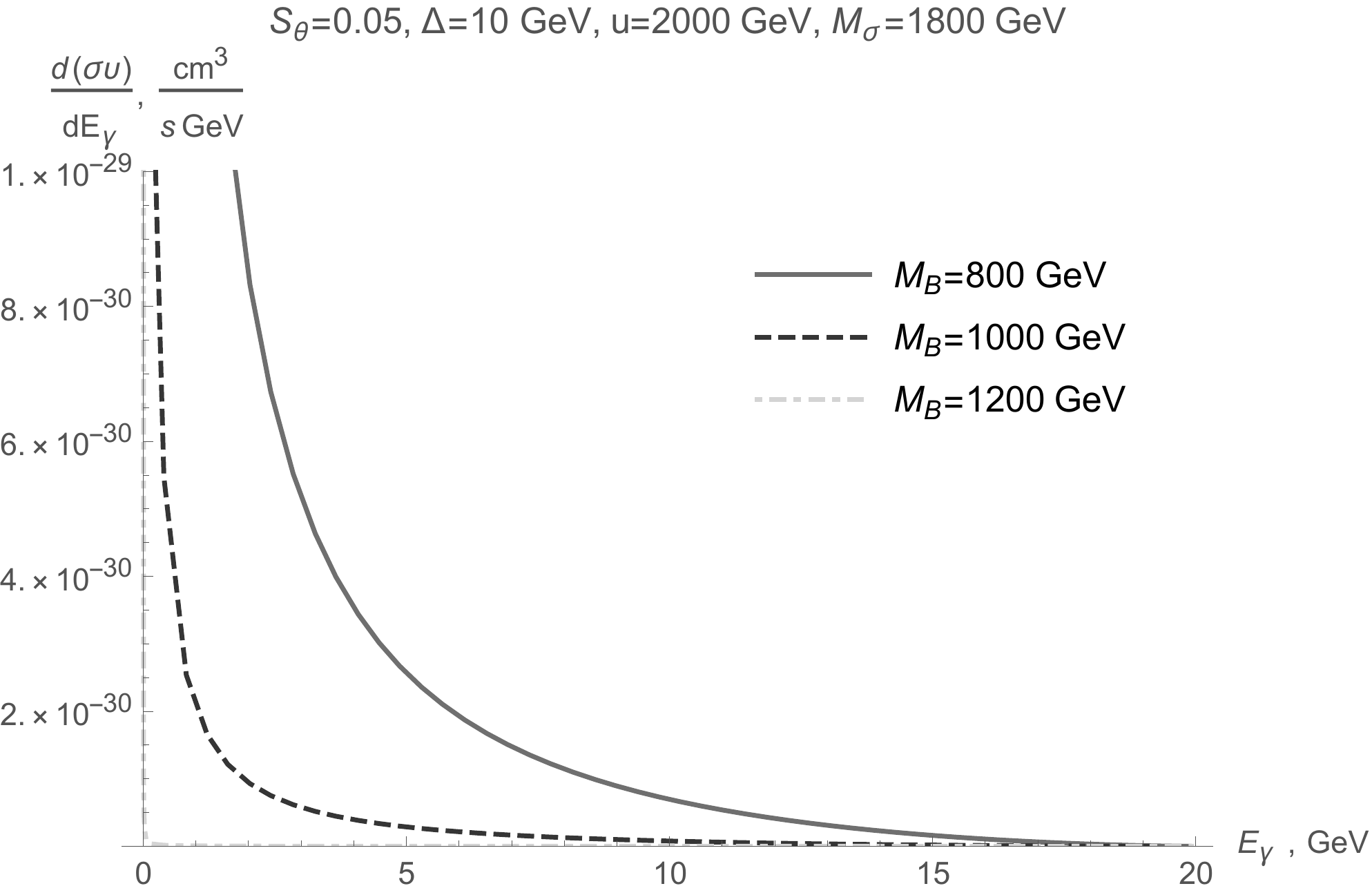}}
	\end{minipage}
	\vfill
	\begin{minipage}[h]{0.5\linewidth}
		\centering{\includegraphics[width=0.7\linewidth]{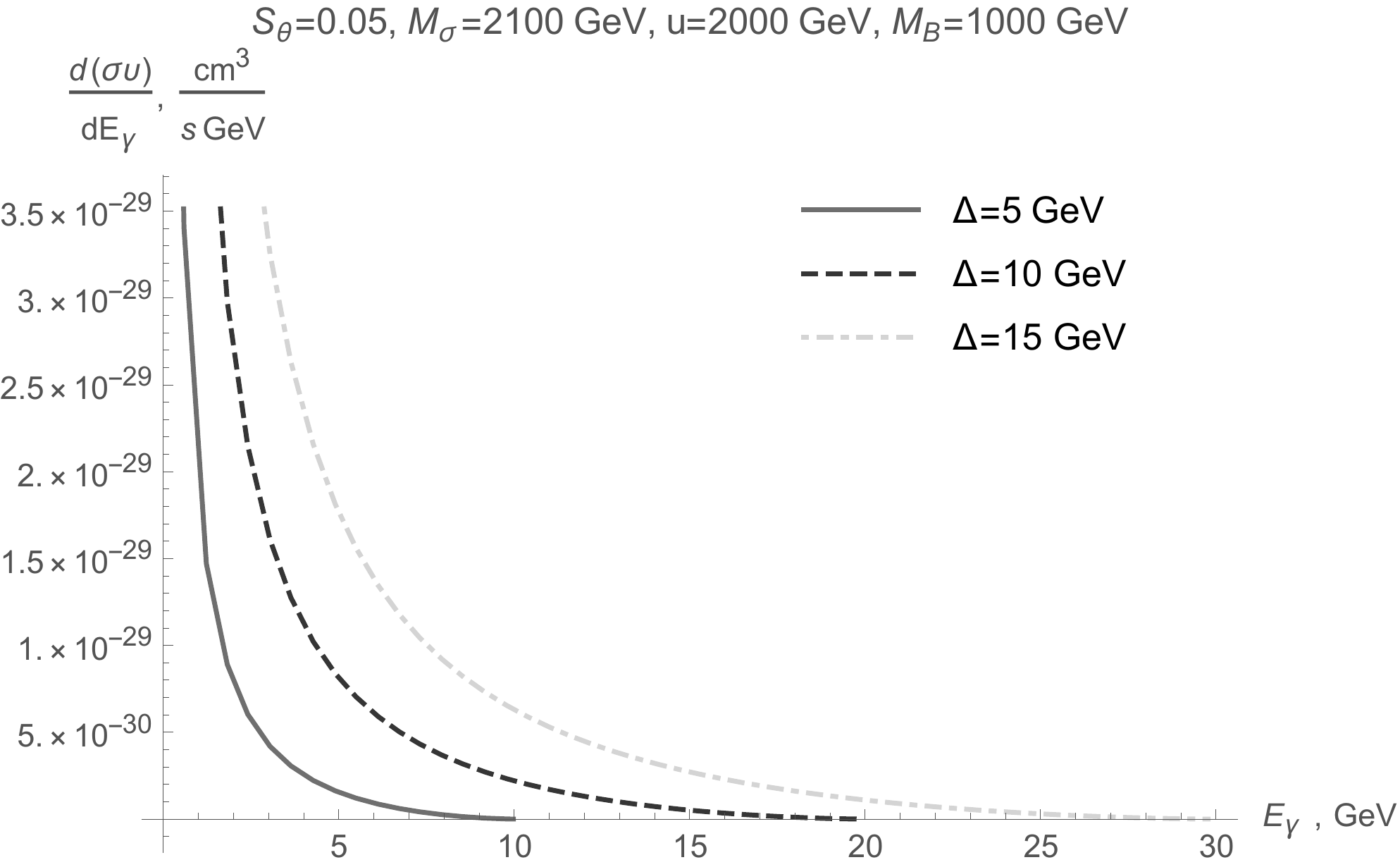}}
	\end{minipage}
	\caption{Differential cross sections for $B^0B^0 \to \tilde\pi^+ \tilde \pi^- \gamma$ process for various model parameters.} 
	\label{spectrum}
\end{figure}

With an integration of the differential cross section from energies $\sim (0.1-0.2)\,\mbox{GeV}$ up to 
$2\cdot \Delta_m$, we have found the total cross section, its values are shown in Fig.5 in dependence on $M_{\tilde \sigma}$ for various mass splittings and masses of $B^0$. Obviously, there are some features of the effect considered.  First, the resonance structure is manifested at $M_{\tilde \sigma}$ due to s-channel contributions, and the total cross section is practically independent on $\tilde \sigma-$ meson mass when its value is 
$\geq (2.0-2.5) \,\mbox{TeV}$. We also estimate total flux of photons using $\sigma_{tot}$ values between $10^{-28}$ and $ 10^{-26}\, cm^3/s$ and the Navarro-Frank-White profile for the dark matter density, $\rho_{NFW} (l,\theta,\phi)$. We used also known astrophysical J-factor for the Galaxy center, namely, we take the angular resolution $\sim 1^{\circ}$ and the value of  $J \approx 10^{-21}$.

\begin{figure}[h]
	\begin{minipage}[h]{0.5\linewidth}
		\centering{\includegraphics[width=0.7\linewidth]{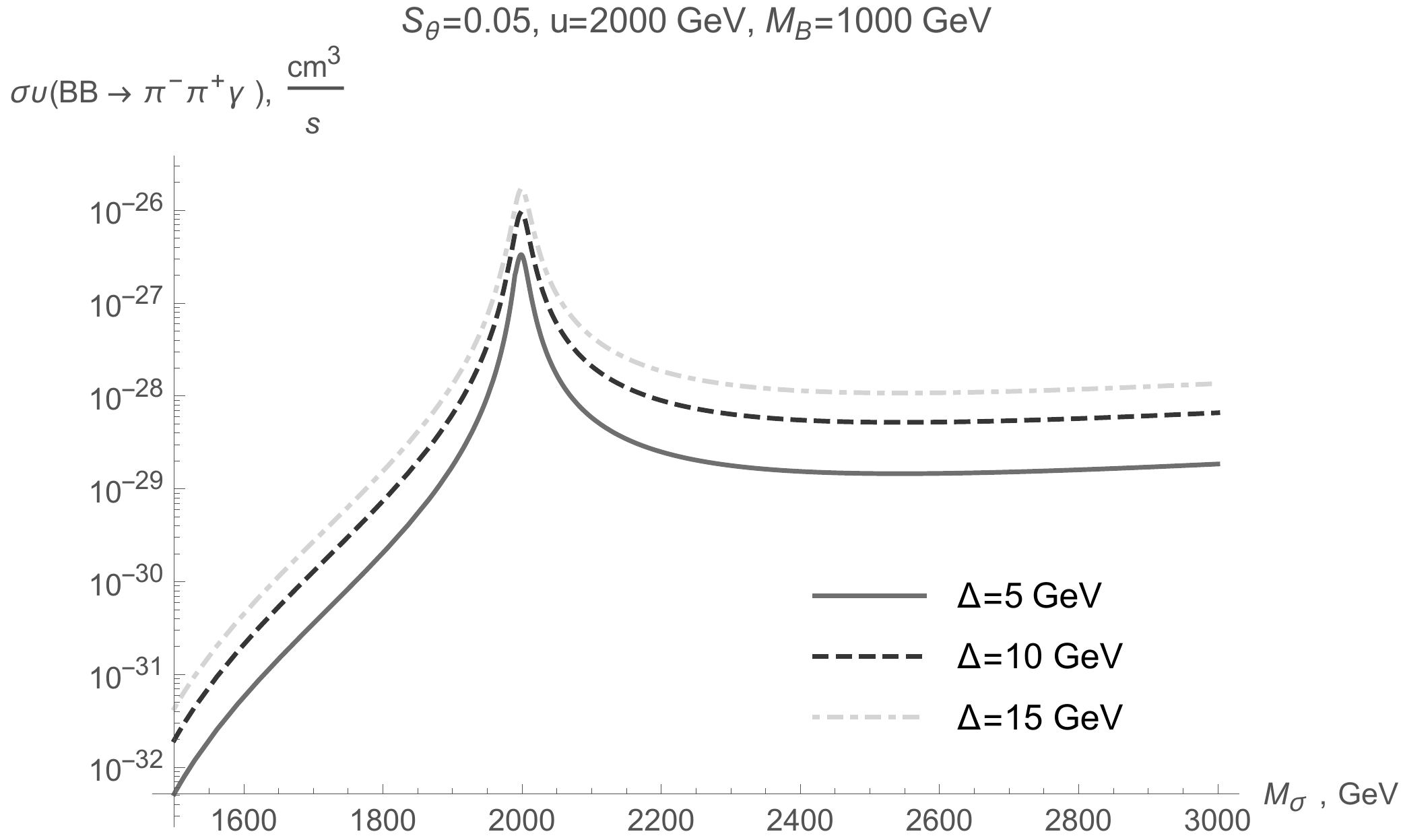}}
	\end{minipage}
	\hfill
	\begin{minipage}[h]{0.5\linewidth}
		\centering{\includegraphics[width=0.7\linewidth]{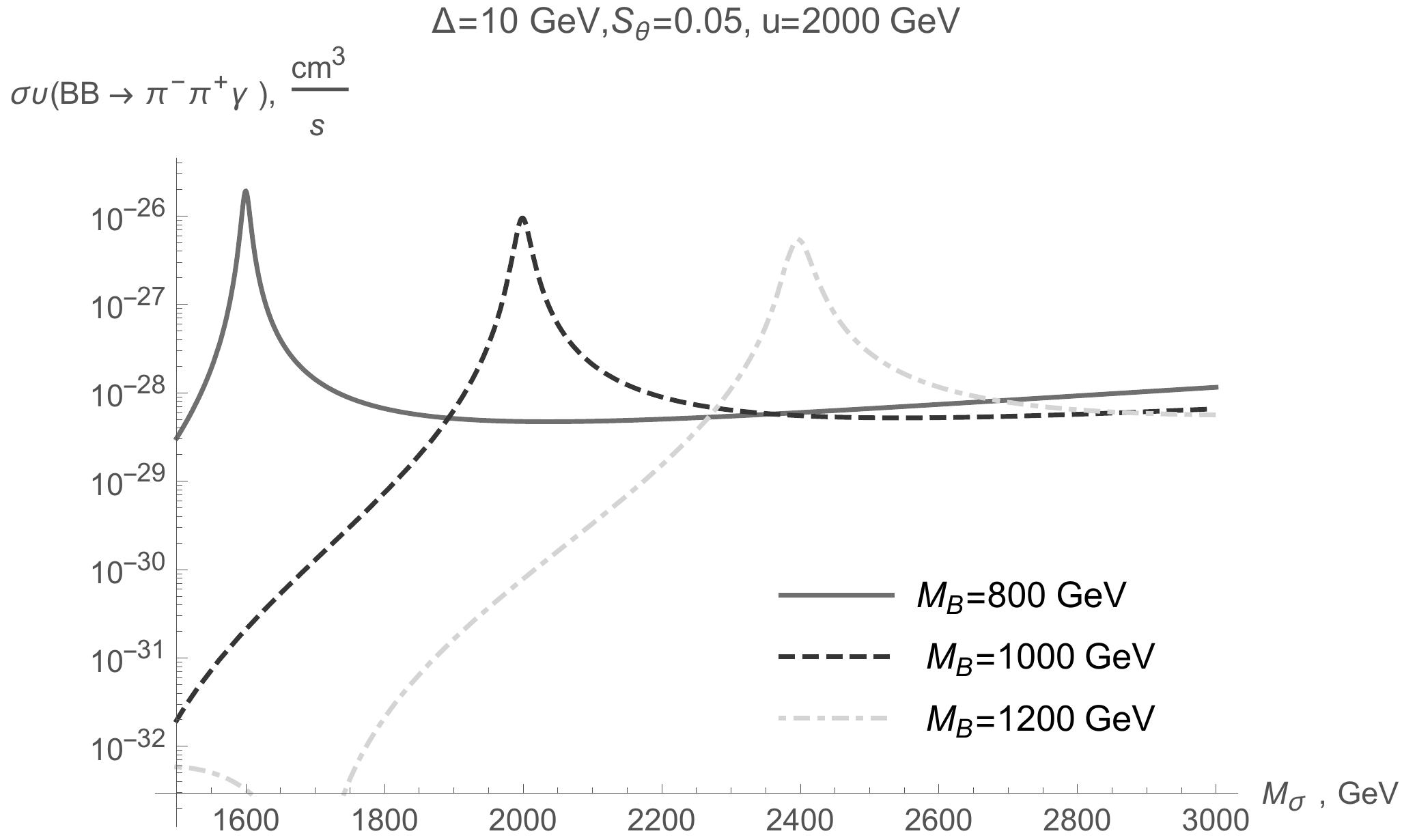}}
	\end{minipage}
	\caption{Total cross section for the process if transition between the DM components with FSR, values of parameters are depicted in figures.}
	\label{Tot}
\end{figure}	

Then, the values of total gamma-flux of low-energy photons produced by transitions between the dark matter components near Galaxy center are the following: 

\begin{equation}\label{Ftot}
\Phi (E_{\gamma}) \approx (0.9 - 1.5)\cdot (10^{-14} - 10^{-12})\, cm^{-2}s^{-1}.
\end{equation}

However, $J$-factor for the Galaxy center can increase up to an order or two if the parameter $\gamma$ in NFW profile changes from 1 to 1.4 to simulate the DM spike in the DM distribution near the GC. Then, the flux also increases up to two orders. 

In this minimal scenario the mass splitting is much lower than the DM masses, so, there arises diffuse photons with energies in a narrow limited area. Certainly, these photons are only an admixture for (monochromatic) radiation from the DM annihilation into photon pairs. This luminescence of the DM is, however, too small to explain the whole excess of GeV photons from GC.

Note also that scenarios with a complex DM sector structure should be analyzed carefully in the case of $\Delta_m \sim M_{DM}$: the DM candidates can be freezed out at different temperatures, so, they can be produced at different stages and contribute separately to features of evolution processes.

Of course, the possible effect of small photonic flux from regions with the increased DM density is specific because it does not lead to the resorption of dense DM clumps. Total mass and the particle number density does not change practically in this process. 
Indeed, there takes places also an "ordinary" annihilation of DM components into two photons or into pairs of SM particles with subsequent photon emission from final or intermediate bosons, leptons and quarks. However, monochromatic photons with energies of the order of the DM masses are separated by an energy gap in the full spectrum of emitted photons. 
 
Unfortunately, a large background is produced by diffuse FSR from the SM particles; the total gamma flux can be noticeably larger than the indicated effect. So, analysis of the photons spectrum at GeV energies is a difficult task.

Indeed, detection and selection of a (nearly constant) photonic component with energies of the order of $(1-10)\, \mbox{GeV}$ can indicate the presence of some structure in the DM mass spectrum, or the possibility of transitions between exited levels in the spectrum of states, as it can occur in the hadronic DM scenario\cite{Kuksa}.

This specific radiation also should be collimated with a some (small) angular aperture if it comes from some “point sources” – GC, dwarf galaxies, subhaloes or other types of DM regions with high density. If the DM clump not very far ($\sim 0.1\, \mbox{pc}$) from our space telescopes, the low-energy limited flux of photons can be seen and recognized. 

Note, inverse case when the neutral H-pions are heavier also should be considered, however, then annihilation into the (lighter) $B^0$-components with radiation of photons difficult - diffuse photons production takes place mostly due to VIB from H-pions and/or H-quarks loops and corresponding cross section should be smaller.

\section{Conclusion}

As some additional considerations, it should be noted that stable DM candidates can be produced from H-quark-gluon plasma at early stages at large temperatures. Besides, due to high scale of H-vacuum condensates, H-hadronization should occur before the QCD hadronization, so the photons from transitions between various H-states can contribute significantly to total density of radiation. This process can maintain the plasma temperature as a kind of delay mechanism that prevents cooling during expansion, in accordance with the Le Chatelier principle.  

This type of annihilation induced by transition between the DM components, is interesting also from the point of view of the DM accumulation inside massive objects – red giants, white dwarfs and the possible dark stars at early stage. In this case, photons, leptons, and neutrinos generated during the transition between components will heat up the interior of the gravitationally coupled system more slowly than the annihilation of DM into SM particles would do (this reaction, of course, also takes place, but with a noticeably smaller cross section for some model parameter values). In the case, the  dark star life time in the relatively “cold” state should icrease.

Moreover, if such reactions dominate, the gravitating mass of the object also will changes slowly. Energies of the photons emitted from such objects will be distributed over significantly different regions separated by a gap of the order of the DM mass. Such an analysis would be reasonable for (early) dark stars with long lifetimes. Their thermonuclear heating is actually replaced by an energy release during the annihilation of DM particles. The discussed effect shows that the presence of a specific complex structure of DM states can be important for the dark stars study. Namely, the luminosity of dark stars can be provided also by low-energy component which is induced by transitions between the DM objects within the stars.

Thus, it was considered the scenario in which one DM component can be effectively go into another through intermediate stage of charged H-pions production and decay. Certainly, it is possible an annihilation of $B^0B^0$ pair into standard quarks and gauge bosons (via scalar exchanges or loops), but it turned out that there is a region of parameters where the cross section for annihilation into hyperpions dominates. Note, the lifetime of charged H-pions is larger than the lifetime of gauge bosons. So, the emission of photons from intermediate charged states, in principle, could be observed. The (small) flux of photons is proportional to the squared DM density, so the most interesting should be to study intensity of such radiation from the GC (from where an increased flux of low-energy photons is observed) or from probable DM clumps. Note also, the DM number density in these processes does not change, intermediate charged H-pions produce neutral stable H-pions together with the low-energy secondaries such as leptons and neutrinos. Besides the low-energy characteristic radiation with small flux, the effect obviously leads to the burning out of heavier DM component transforming it to another one. This process, however, is very slow, since the DM concentration is low. Both DM components are practically in equilibrium, so that a small change in their concentrations is hardly noticeable.

It would be interesting to analyze another scenarios of the SM extensions with a complex structure of DM sector and rich phenomenology to study possible observed manifestations of multicomponent dark matter using multi-messenger approach for the analysis of possible signals.

\section*{Acknowledgements}
The work was supported by grant of the Russian Science Foundation (Project No-18-12-00213-P)


\end{document}